# FWM in One-dimensional Nonlinear Photonic Crystal and Theoretical Investigation of Parametric Down Conversion (Steady State Analysis)

**A. Rostami and M. Boozarjmehr**

Photonics and Nanocrystals Research Lab., (PNRL), Faculty of Electrical and Computer Engineering, University of Tabriz, Tabriz 51664, Iran
Tel/Fax: +98 411 3393724
E-mail: rostami@tabrizu.ac.ir

**Abstract-** The light propagation through one-dimensional photonic crystal using Four-wave mixing (FWM) nonlinear process is modeled. The linear and nonlinear indexes of refraction are approximated with the first Fourier harmonic term. Based on this approximation, a complete set of coupled wave equations, including pump fields depletion, for description of FWM process and conversion efficiency from pump to signal and idler waves for periodic structures is presented. The derived coupled wave equations are evaluated numerically. Some of important system parameters effects on FWM performance are investigated. The obtained relations are suitable and can easily be applied for description of Wavelength Division Multiplexing (WDM) optical signals transmitted through $\chi^{(3)}$ (parametric process) nonlinear fiber Bragg Gratings compatible to optical fiber communications.

**I. Introduction-** Nonlinear phenomenon in optical range plays critical role in all-optical networks. For example, Kerr-like nonlinear medium is very interesting for realization of full-optical devices and systems, such as optical limiter, optical switch, optical A/D and D/A, optical Multiplex and De-multiplex, and many other interesting applications [1]. As a second category of application, generation of entangled photon pairs has been considered as an important topic, because of its key role in realizing quantum communication technology, including quantum cryptography, quantum dense coding, and quantum teleportation [2-4]. Anther important application is wave mixing in optical range [5, 6].

All of these advantages and other nonlinear medium based applications are so interesting from practical point of view, especially for obtaining all-optical communication network.

One of nonlinear important phenomenon is Four-wave Mixing (FWM) process. FWM has long been studied especially in the optical fibers and in investigation of the wavelength and dense wavelength division multiplexing (WDM and DWDM) systems. FWM is really a photon-photon scattering process, during which two photons from a relatively high-intensity beam, called pump beam, scatter through third-order ($\chi^{(3)}$) nonlinearity of a material to generate two correlated photons, called signal and idler photons respectively [7-9].

In homogeneous nonlinear media (such as bulk material), efficient exchange of energy between interacting modes of the electromagnetic field is determined by the linear and nonlinear susceptibilities of the medium. So, successful achievement of the proposed applications strongly depends on the nonlinearity strength and medium structure. But these materials suffer from several problems which some of them are mentioned below:

1. $\chi^{(3)}$ Nonlinearity is usually small compared to $\chi^{(2)}$.
2. Wavelengths of signal and idler photons are close to pump wavelengths.
3. In the case of small conversion efficiency, even small amounts of pump beam scattering generates large background count rates that mask the detection of correlations between signal and idler photons. (on the other hand scattering of the pump fields tends to mask the desired quantum effects).

Many of the problems associated with $\chi^{(3)}$ nonlinear optical materials can be eliminated by using suitable structures such as single mode optical fibers. These optical fibers have extremely low loss, small confinement cross section and can be as long as several kilometers. The nonlinearity is an off-resonance $\chi^{(3)}$ Kerr effect with an ultra fast frequency response extending from dc to well above 10 THz. Although weak, it can give rise to very large nonlinear effect in long fibers.
It is obvious that the material's permittivity determines how phase matched is a given parametric process, whereas the actual coupling of energy between the modes is a function of the material's nonlinear polarizability. In an attempt to circumvent material constraints (second alternative), much works have been focused on the possibility of using periodic media to mediate nonlinear processes. Some of basic important works proposed the introduction of periodic structure into the linear and nonlinear material properties to aid in phase matching parametric interactions [10-15].
The introduction of the periodic nonlinear modulation leads to both flexibility in phase matching and also makes accessible a material's largest nonlinear coefficient. It has been shown that periodic modulation of a nonlinear material's refractive index can lead to enhanced conversion efficiencies in parametric processes.
Photonic crystals were first conceived by John and Yablonovitch [16,17], and have been widely used in all fields of optics, so especial arrangement of linear and nonlinear index of refraction can help us to modify the conversion efficiency of FWM process.
In this paper, we propose a complete set of coupled wave equations describing FWM in one-dimensional nonlinear photonic crystal for the first time. The derived relations include all system parameters and input status. Our consideration concentrate on $1.55\mu m$ which is interesting for optical communication. Also, the proposed periodic structure can be imagined as nonlinear fiber Bragg Grating. After derivation of the coupled wave equations for all field components (forward and reflected components), numerical methods have been used for simulation of the process. Simulation results for conversion efficiency are presented both for co-propagating and counter-propagating signal fields. Also, we try to enhance conversion efficiencies using FWM process. Our obtained results show that by increasing the number of the medium layers, conversion efficiency increases.
This paper is organized as follows,
Mathematical formulation and coupled wave equations are discussed in section II, where the mathematical derivation of the proposed system is presented. Simulation and numerical evaluation of the obtained equations is investigated in section III. Finally the paper ends with a conclusion.

**II. Mathematical Modeling-** In this section, we try to present a mathematical model for description of FWM in one-dimensional nonlinear photonic crystal. Typical index of refraction profiles for the considered system are illustrated in Fig. (2-1).

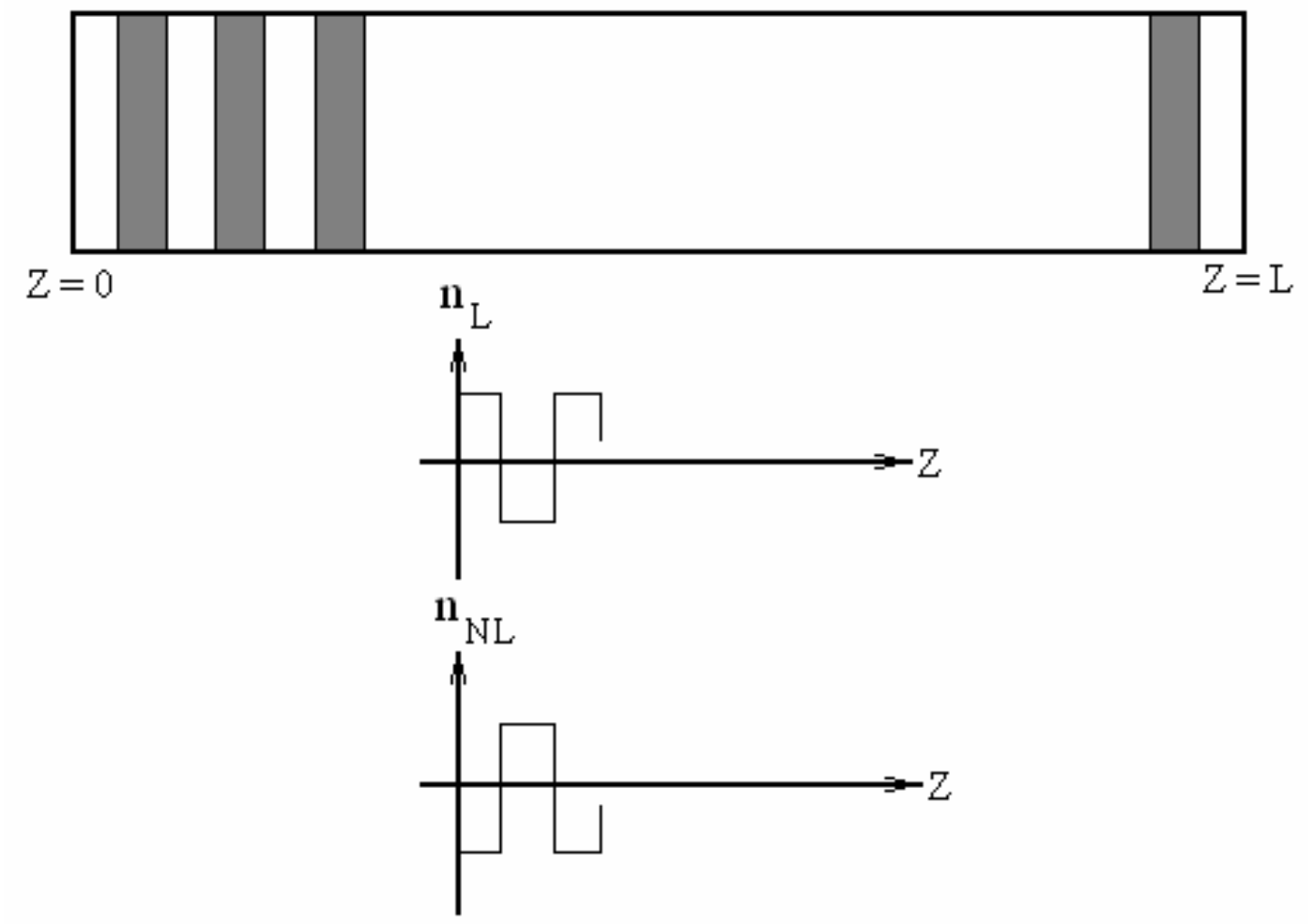

Fig. (2-1): One-dimensional nonlinear photonic crystal and the indexes of refraction distribution

For the proposed structure the index of refraction is given as follows:

$$n = n_0 + a_1 \cos(2k_0 + \delta)z + a_2 |E|^2 \cos(2k_0 + \delta)z, \tag{2-1}$$

where $n_0$, $a_1$, $k_0$, $\delta$, $a_2$ and $E$ are the average index of refraction, the first harmonic coefficient of Fourier expansion for the linear index of refraction, average incident wave vector, phase mismatch between incident wave vector and periodic structure's wave vector, the first harmonic coefficient of Fourier expansion for the nonlinear index of refraction and the applied electric field, respectively.
We have assumed the following field distribution for FWM process in the periodic structure,

$$\begin{aligned} E = {} & (E_{+1} e^{ik_1 z} + E_{-1} e^{-ik_1 z}) e^{-i\omega_1 t} + (E_{+2} e^{ik_2 z} + E_{-2} e^{-ik_2 z}) e^{-i\omega_2 t} + \\ & (E_{+3} e^{-ik_3 z} + E_{-3} e^{ik_3 z}) e^{-i\omega_3 t} + (E_{+4} e^{-ik_4 z} + E_{-4} e^{ik_4 z}) e^{-i\omega_4 t} + c.c., \end{aligned} \tag{2-2}$$

where $E_{\pm i}$, $k_i$, and $\omega_i$ are amplitudes of the forward and backward pump , signal and idler fields, their wave vectors and frequencies for all components, respectively. Also the following relation stands for phase mismatching condition between four wave vectors should be satisfied,

$$\Delta k = k_1 + k_2 - k_3 - k_4 \tag{2-3}$$

For the proposed nonlinear medium, nonlinear polarization is [6],

$$P_{NL} = \varepsilon_0 [A(z)(E.E^*)E + \frac{1}{2} B(z)(E.E)E^*] + c.c., \tag{2-4}$$

where $A(z)$ and $B(z)$ are nonlinear polarization constants related to nonlinear medium distribution profile and are given as follows,

$$A(z) = B(z) = -A[e^{i(2k_0+\delta)z} + e^{-i(2k_0+\delta)z}], \tag{2-5}$$

where

$$k_0 = \frac{k_1 + k_2 + k_3 + k_4}{4}. \tag{2-6}$$

Now, for obtaining the coupled wave equations, the electric filed and the nonlinear polarization should satisfy the Maxwell's wave equation,

$$\frac{\partial^2 E}{\partial z^2} - \frac{n^2}{c^2} \frac{\partial^2 E}{\partial t^2} = \mu_0 \frac{\partial^2 P_{NL}}{\partial t^2}, \tag{2-7}$$

where $n$, $c$ and $\mu_0$ are the index of refraction, speed of light in free space and medium permeability, respectively.
Because of small perturbation in the index of refraction, the following approximation is used for index of refraction appeared in Maxwell's wave equation,

$$n^2 = n_0^{\ 2} + n_0 a_1 (e^{i(2k_0+\delta)z} + e^{-i(2k_0+\delta)z}) - n_0 a_2 (e^{i(2k_0+\delta)z} + e^{-i(2k_0+\delta)z}) |E|^2 \tag{2-8}$$

Finally, after substitution (2-2), (2-4) and (2-8) in Eq. (2-7) and doing some mathematical simplifications, the following coupled wave equations are obtained:

$$\frac{\partial E_{+1}}{\partial z}=\frac{i\omega_1^2 n_0 a_1 E_{-1}e^{i\delta z}}{2k_1c^2}-\frac{i\omega_1^2(n_0a_2+\frac{3}{4}A)}{k_1c^2}\times$$
$$\{\alpha_1E_{-1}e^{i\delta z}+\alpha_2E_{+1}e^{i\delta z}+\alpha_3E_{+1}e^{-i\delta z}+2E_{-3}^*E_{-4}^*E_{+2}e^{i\delta z}e^{i\Delta kz}\}, \qquad (2\text{-}9)$$

where

$$\alpha_1=2|E_{+1}|^2+2|E_{+2}|^2+2|E_{+3}|^2+2|E_{+4}|^2+|E_{-1}|^2+2|E_{-2}|^2+2|E_{-3}|^2+2|E_{-4}|^2,$$
$$\alpha_2=2E_{+3}E_{-3}^*+2E_{+4}E_{-4}^*+2E_{-2}E_{+2}^*, \qquad (2\text{-}10)$$
$$\alpha_3=E_{+1}E_{-1}^*+2E_{+2}E_{-2}^*+2E_{-3}E_{+3}^*+2E_{-4}E_{+4}^*.$$

Eq. (2-9) illustrates the coupled wave equation for the pump field propagating from left to right in the medium.

$$-\frac{\partial E_{-1}}{\partial z}=\frac{i\omega_1^2 n_0 a_1 E_{+1}e^{-i\delta z}}{2k_1c^2}-\frac{i\omega_1^2(n_0a_2+\frac{3}{4}A)}{k_1c^2}\times$$
$$\{\Gamma_1E_{+1}e^{-i\delta z}+\Gamma_2E_{-1}e^{-i\delta z}+\Gamma_3E_{-1}e^{i\delta z}+2E_{-3}E_{-4}E_{+2}^*e^{-i\delta z}e^{-i\Delta kz}\}, \qquad (2\text{-}11)$$

where

$$\Gamma_1=|E_{+1}|^2+2|E_{+2}|^2+2|E_{+3}|^2+2|E_{+4}|^2+2|E_{-1}|^2+2|E_{-2}|^2+2|E_{-3}|^2+2|E_{-4}|^2,$$
$$\Gamma_2=2E_{+2}E_{-2}^*+2E_{-3}E_{+3}^*+2E_{-4}E_{+4}^*, \qquad (2\text{-}12)$$
$$\Gamma_3=E_{-1}E_{+1}^*+2E_{-2}E_{+2}^*+2E_{+3}E_{-3}^*+2E_{+4}E_{-4}^*.$$

Eq. (2-11) illustrates the coupled wave equation for the reflected part of the pump field propagating from right to left in the medium.

$$\frac{\partial E_{+2}}{\partial z}=\frac{i\omega_2^2 n_0 a_1 E_{-2}e^{i\delta z}}{2k_2c^2}-\frac{i\omega_2^2(n_0a_2+\frac{3}{4}A)}{k_2c^2}\times$$
$$\{\beta_1E_{-2}e^{i\delta z}+\beta_2E_{+2}e^{i\delta z}+\beta_3E_{+3}e^{-i\delta z}+2E_{-3}^*E_{-4}^*E_{+1}e^{i\delta z}e^{i\Delta kz}\}, \qquad (2\text{-}13)$$

where

$$\beta_1=2|E_{+1}|^2+2|E_{+2}|^2+2|E_{+3}|^2+2|E_{+4}|^2+2|E_{-1}|^2+|E_{-2}|^2+2|E_{-3}|^2+2|E_{-4}|^2,$$
$$\beta_2=2E_{+3}E_{-3}^*+2E_{+4}E_{-4}^*+2E_{-1}E_{+1}^*, \qquad (2\text{-}14)$$
$$\beta_3=2E_{+1}E_{-1}^*+E_{+2}E_{-2}^*+2E_{-3}E_{+3}^*+2E_{-4}E_{+4}^*.$$

Eq. (2-13) illustrates the coupled wave equation for the second pump field propagating from left to right in the medium.

$$-\frac{\partial E_{-2}}{\partial z}=\frac{i\omega_2^2 n_0 a_1 E_{+2}e^{-i\delta z}}{2k_2c^2}-\frac{i\omega_2^2(n_0a_2+\frac{3}{4}A)}{k_2c^2}\times$$
$$\{\theta_1E_{+2}e^{-i\delta z}+\theta_2E_{-2}e^{-i\delta z}+\theta_3E_{-3}e^{i\delta z}+2E_{-3}E_{-4}E_{+1}^*e^{-i\delta z}e^{-i\Delta kz}\}, \qquad (2\text{-}15)$$

where

$$\theta_1 = 2|E_{+1}|^2 + |E_{+2}|^2 + 2|E_{+3}|^2 + 2|E_{+4}|^2 + 2|E_{-1}|^2 + 2|E_{-2}|^2 + 2|E_{-3}|^2 + 2|E_{-4}|^2,$$
$$\theta_2 = 2E_{+2}E_{-2}^* + 2E_{-3}E_{+3}^* + 2E_{-4}E_{+4}^*, \qquad (2\text{-}16)$$
$$\theta_3 = 2E_{-1}E_{+1}^* + E_{-2}E_{+2}^* + 2E_{+3}E_{-3}^* + 2E_{+4}E_{-4}^*.$$

Eq. (2-15) illustrates the coupled wave equation for the reflected part of the second pump field propagating from right to left in the medium.

$$-\frac{\partial E_{+3}}{\partial z} = \frac{i\omega_3^2 n_0 a_1 E_{-3} e^{-i\delta z}}{2k_3c^2} - \frac{i\omega_3^2 (n_0 a_2 + \frac{3}{4}A)}{k_3 c^2} \times \qquad (2\text{-}17)$$
$$\{\gamma_1 E_{-3} e^{-i\delta z} + \gamma_2 E_{+3} e^{-i\delta z} + \gamma_3 E_{+3} e^{i\delta z} + 2E_{+1}E_{+2}E_{-4}^* e^{-i\delta z} e^{i\Delta kz}\},$$

where

$$\gamma_1 = 2|E_{+1}|^2 + 2|E_{+2}|^2 + 2|E_{+3}|^2 + 2|E_{+4}|^2 + 2|E_{-1}|^2 + 2|E_{-2}|^2 + |E_{-3}|^2 + 2|E_{-4}|^2,$$
$$\gamma_2 = 2E_{+1}E_{-1}^* + 2E_{+2}E_{-2}^* + 2E_{-4}E_{+4}^*, \qquad (2\text{-}18)$$
$$\gamma_3 = 2E_{-1}E_{+1}^* + 2E_{-2}E_{+2}^* + E_{+3}E_{-3}^* + 2E_{+4}E_{-4}^*.$$

Eq. (2-17) illustrates the coupled wave equation for the signal field propagating from right to left in the medium.

$$\frac{\partial E_{-3}}{\partial z} = \frac{i\omega_3^2 n_0 a_1 E_{+3} e^{i\delta z}}{2k_3c^2} - \frac{i\omega_3^2 (n_0 a_2 + \frac{3}{4}A)}{k_3 c^2} \times \qquad (2\text{-}19)$$
$$\{\psi_1 E_{+3} e^{i\delta z} + \psi_2 E_{-3} e^{i\delta z} + \psi_3 E_{-3} e^{-i\delta z} + 2E_{-4}E_{+1}^* E_{+2}^* e^{i\delta z} e^{-i\Delta kz}\},$$

where

$$\psi_1 = 2|E_{+1}|^2 + 2|E_{+2}|^2 + |E_{+3}|^2 + 2|E_{+4}|^2 + 2|E_{-1}|^2 + 2|E_{-2}|^2 + 2|E_{-3}|^2 + 2|E_{-4}|^2,$$
$$\psi_2 = 2E_{+4}E_{-4}^* + 2E_{-2}E_{+2}^* + 2E_{-1}E_{+1}^*, \qquad (2\text{-}20)$$
$$\psi_3 = 2E_{+1}E_{-1}^* + 2E_{+2}E_{-2}^* + 2E_{-3}E_{+3}^* + 2E_{-4}E_{+4}^*.$$

Eq. (2-19) illustrates the coupled wave equation for the reflected part of the signal field propagating from left to right in the medium.

$$-\frac{\partial E_{+4}}{\partial z} = \frac{i\omega_4^2 n_0 a_1 E_{-4} e^{-i\delta z}}{2k_4c^2} - \frac{i\omega_4^2 (n_0 a_2 + \frac{3}{4}A)}{k_4 c^2} \times \qquad (2\text{-}21)$$
$$\{\eta_1 E_{-4} e^{-i\delta z} + \eta_2 E_{+4} e^{-i\delta z} + \eta_3 E_{+4} e^{i\delta z} + 2E_{+1}E_{+2}E_{-3}^* e^{-i\delta z} e^{i\Delta kz}\},$$

where

$$
\begin{aligned}
\eta_1 &= 2|E_{+1}|^2 + 2|E_{+2}|^2 + 2|E_{+3}|^2 + 2|E_{+4}|^2 + 2|E_{-1}|^2 + 2|E_{-2}|^2 + 2|E_{-3}|^2 + |E_{-4}|^2, \\
\eta_2 &= 2E_{+1}E_{-1}^* + 2E_{+2}E_{-2}^* + 2E_{-3}E_{+3}^*, \qquad (2\text{-}22) \\
\eta_3 &= 2E_{-1}E_{+1}^* + 2E_{-2}E_{+2}^* + 2E_{+3}E_{-3}^* + E_{+4}E_{-4}^*.
\end{aligned}
$$

Eq. (2-21) illustrates the coupled wave equation for the idler field propagating from right to left in the medium.

$$
\begin{aligned}
\frac{\partial E_{-4}}{\partial z} &= \frac{i\omega_4^2 n_0 a_1 E_{+4} e^{i\delta z}}{2k_4 c^2} - \frac{i\omega_4^2 (n_0 a_2 + \frac{3}{4} A)}{k_4 c^2} \times \qquad (2\text{-}23) \\
&\{\mathrm{K}_1 E_{+4} e^{i\delta z} + \mathrm{K}_2 E_{-4} e^{i\delta z} + \mathrm{K}_3 E_{-4} e^{-i\delta z} + 2E_{-3}E_{+1}^* E_{+2}^* e^{i\delta z} e^{-i\Delta kz}\},
\end{aligned}
$$

where

$$
\begin{aligned}
\mathrm{K}_1 &= 2|E_{+1}|^2 + 2|E_{+2}|^2 + 2|E_{+3}|^2 + |E_{+4}|^2 + 2|E_{-1}|^2 + 2|E_{-2}|^2 + 2|E_{-3}|^2 + 2|E_{-4}|^2, \\
\mathrm{K}_2 &= 2E_{+3}E_{-3}^* + 2E_{-2}E_{+2}^* + 2E_{-1}E_{+1}^*, \qquad (2\text{-}24) \\
\mathrm{K}_3 &= 2E_{+1}E_{-1}^* + 2E_{+2}E_{-2}^* + 2E_{-3}E_{+3}^* + 2E_{-4}E_{+4}^*.
\end{aligned}
$$

Eq. (2-23) illustrates the coupled wave equation for the reflected part of the idler field propagating from left to right in the medium.

We have solved these equations numerically and the effect of system parameters on conversion efficiency (pump to signal and idler) is considered and illustrated in the next section. The conversion efficiency for forward traveling signal components can be defined as follows,

$$
\eta_+ = \frac{P_{+3}(0)}{P_{+1}(0)}, \qquad (2\text{-}25)
$$

We will call this efficiency as co-propagation efficiency in the following sections. $P_{+3}(0)$ and $P_{+1}(0)$ stands for the forward propagating pump and signal power respectively, which signal and pump fields are applied from right hand side and left hand side to our system respectively. Also, we define conversion efficiency parameter for backward signal power as follows,

$$
\eta_- = \frac{P_{-3}(L)}{P_{+1}(0)}, \qquad (2\text{-}26)
$$

where $P_{-3}(L)$ is backward signal power at the right hand side of crystal.

**III. Simulation Results and Discussion-** In this section some of numerical simulated diagrams are presented for description of FWM efficiency which increases due to the periodic structure based on the derived equations in the previous section. For realization of this subject, we have simulated the equations both for co-propagation and counter-propagation efficiencies. Our simulations show that, the co-propagation conversion efficiency is considerably larger than the counter-propagation case. This subject can be described based on the basic principles of energy transfer between forward and backward propagating modes in Bragg Gratings.
Effect of number of periods on co-propagation conversion efficiency is illustrated in Fig. (1). Conversion efficiency is decreased with decreasing of the nonlinear index of refraction coefficient.

FWM phenomenon is strongly affected by the nonlinear index of refraction coefficient and this effect has been shown in our numerical simulations.
Also, since the conversion efficiency is nonlinearly dependant on the number of periods, it is decreased when we choose much more layers than 600 layers, so it is necessary to keep in mind this subject in design of a photonic crystal to avoid decreasing of the conversion efficiency.

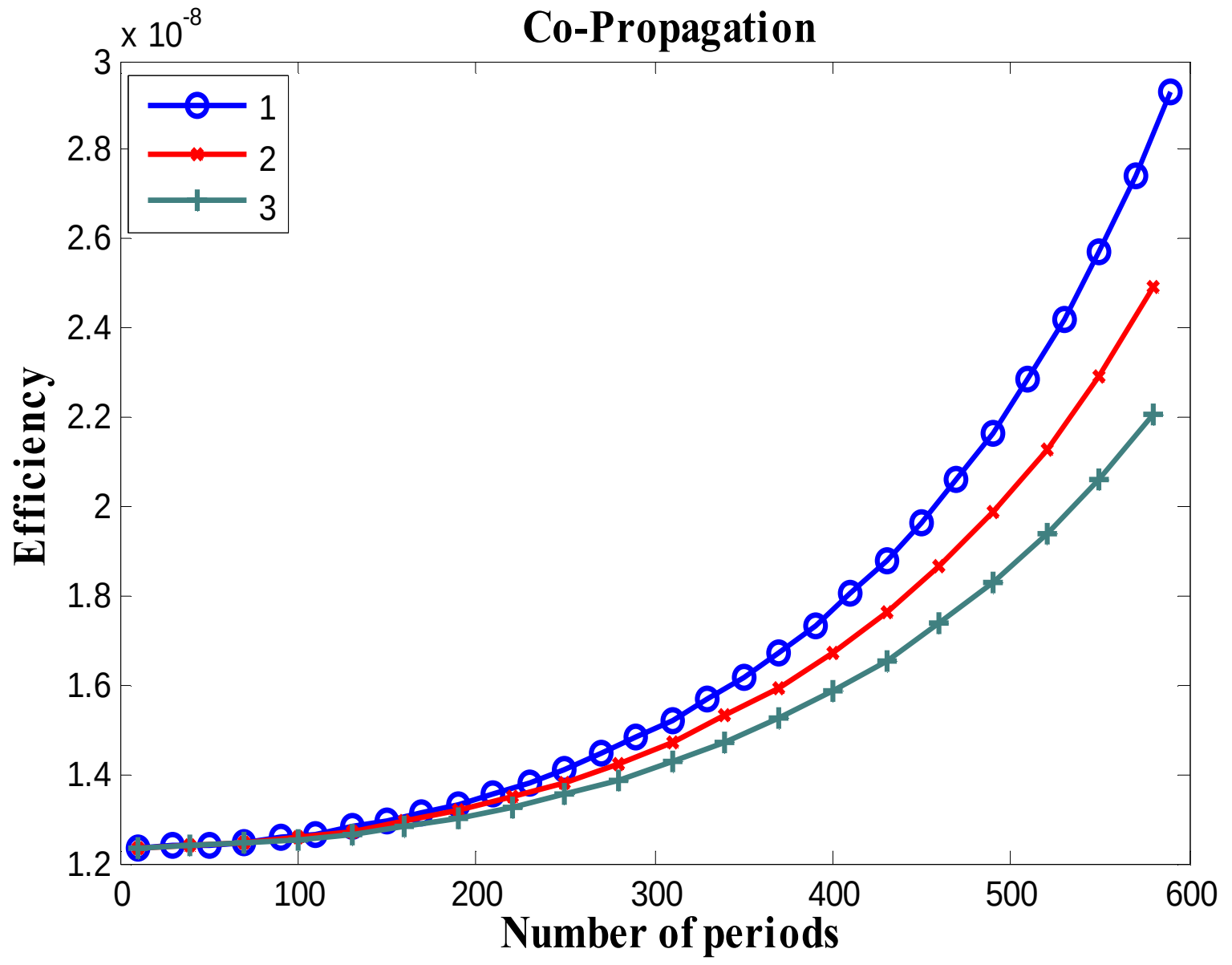

Fig. (1): Efficiency Vs. Number of Periods
$1)\,a_2 = -10^{-15}, 2)\,a_2 = -2\times10^{-15}, 3)\,a_2 = -3\times10^{-15}, n_0 = 3.45, \Delta k = 0, a_1 = 0.001, \delta = 0$

The effect of mismatching between four optical field wave vectors (two pumps, signal and idler fields) on conversion efficiency is illustrated in Fig. (2). by increasing of phase mismatching, conversion efficiency is decreased. This is related to weak energy transfer between propagating modes in this case.

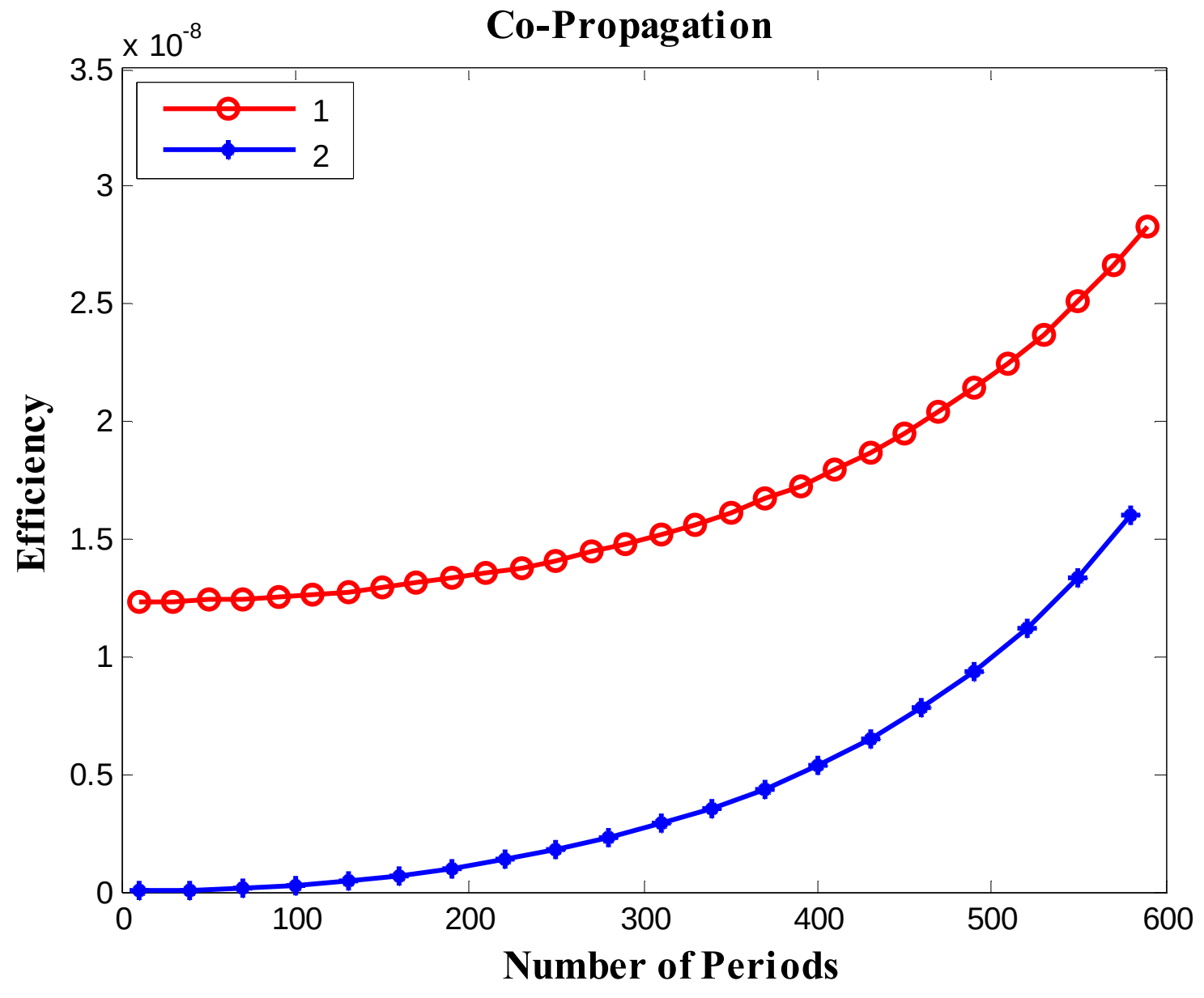

Fig. (2): Efficiency Vs. Number of Periods
$1)\,\Delta k = 0.0001k_0, 2)\,\Delta k = 0.001k_0, 3)\,a_2 = -10^{-15}, n_0 = 3.45, \delta = 0, a_1 = 0.001$

The effect of mismatching between the average wave vector of optical fields and Grating wave vector is demonstrated in Fig. (3). by increasing of mismatching, conversion efficiency decreases and also because of nonlinear relation between conversion efficiency and number of periods, as number of periods becomes larger, the difference between efficiencies in different mismatchings, becomes much more apparent. Also, as we know in Grating structure high efficiency is accessible with large number of periods and this effect shows that for obtaining higher efficiencies, mismatching must be considered as small as possible.

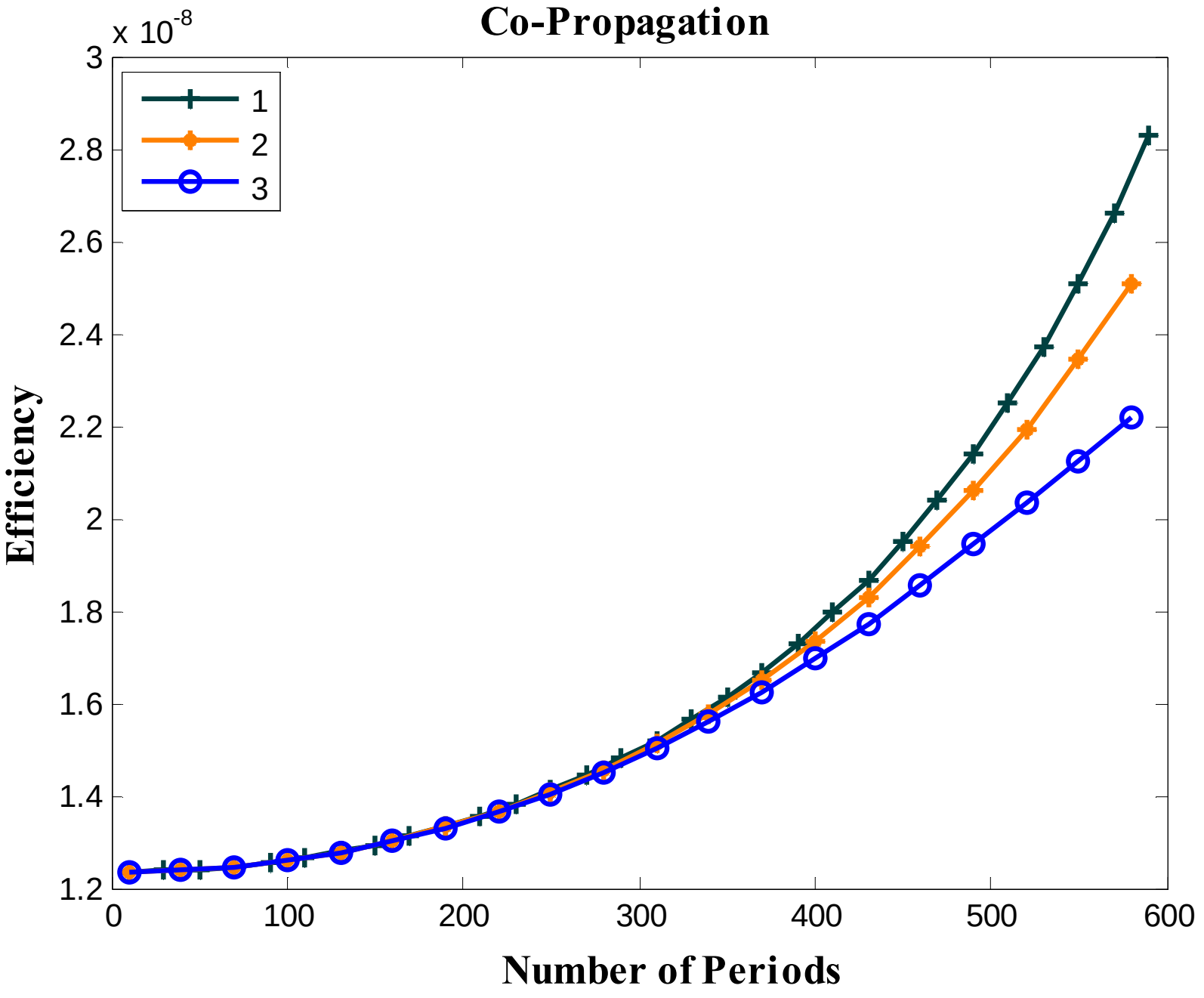

Fig. (3): Efficiency Vs. Number of Periods

$1)\,\delta = 0.001k_0, 2)\,\delta = 0.002k_0, 3)\,\delta = 0.003k_0, n_0 = 3.45, \Delta k = 0, a_1 = 0.001, a_2 = -10^{-15}$

The effect of phase mismatching between Grating wave vector and the average wave vector of four optical fields on conversion efficiency in broad ranges is shown in Fig. (4).

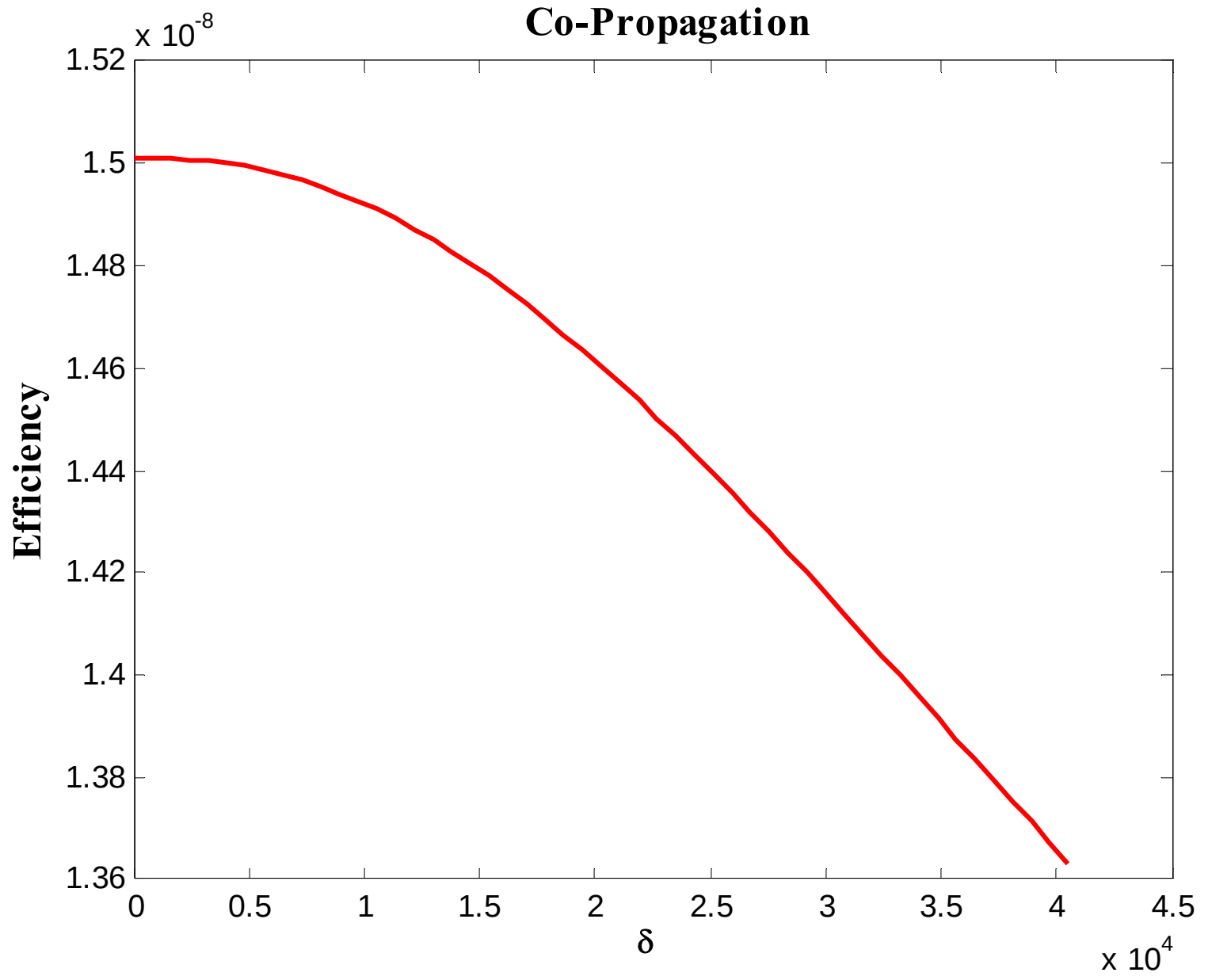

Fig. (4): Efficiency Vs. Phase mismatching between Medium and average applied fields wave vectors

$n_0 = 3.45, \Delta k = 0, a_1 = 0.001, a_2 = -10^{-15}, N = 300$

Also, the effect of the nonlinear index of refraction on the conversion efficiency is shown in Fig. (5).

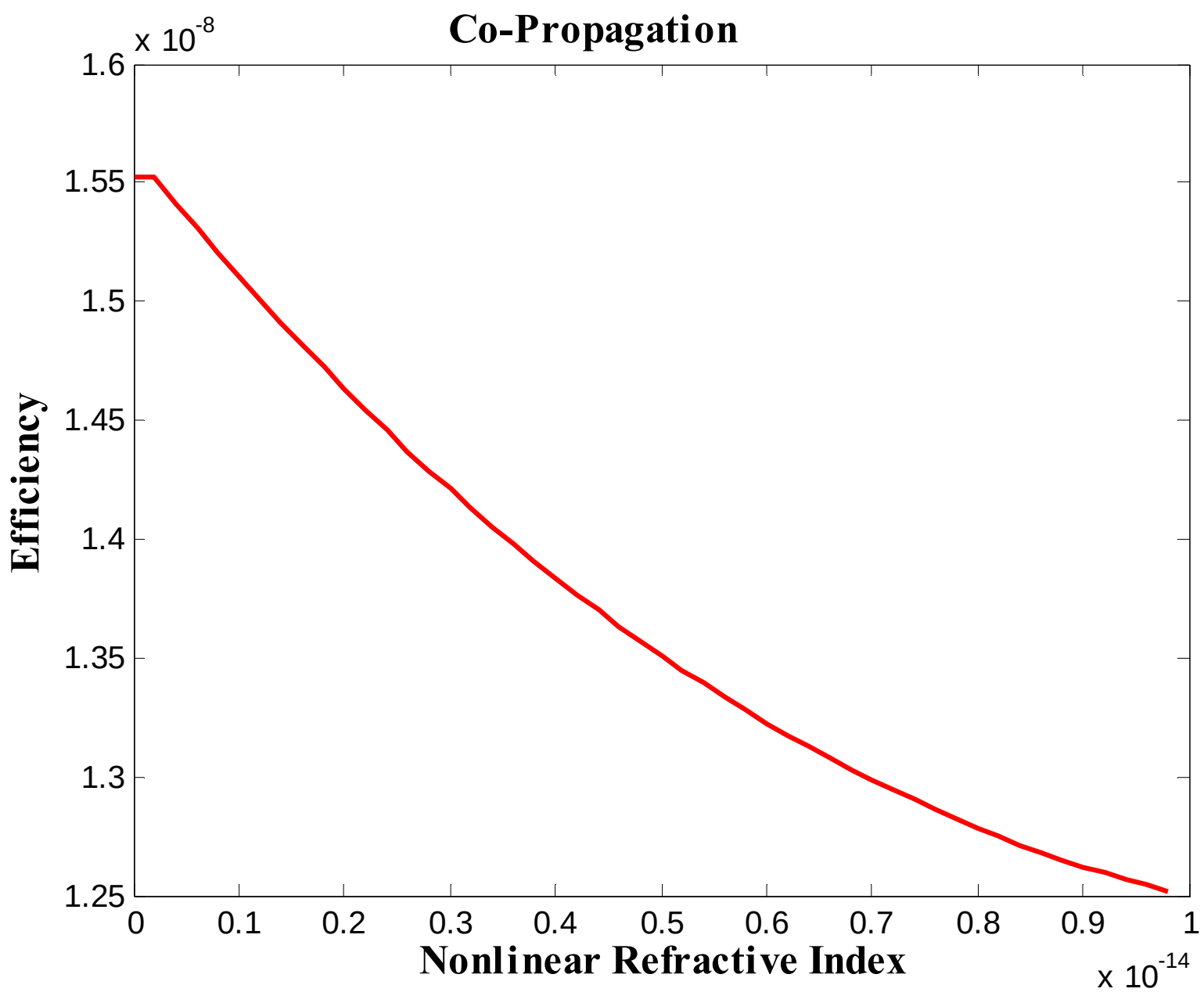

Fig. (5): Efficiency Vs. Absolute Value of Nonlinear Index of Refraction

$n_0 = 3.45, \Delta k = 0, a_1 = 0.001, \delta = 0, N = 300$

The same simulations have been illustrated for counter-propagation case. It should be mentioned that the presented results are not the optimum ones, but they are only typical simulations for some given parameters.

The following simulations belong to counter-propagation case.

The effect of phase mismatching between medium and average applied wave vectors on the conversion efficiency is illustrated in Fig. (6).

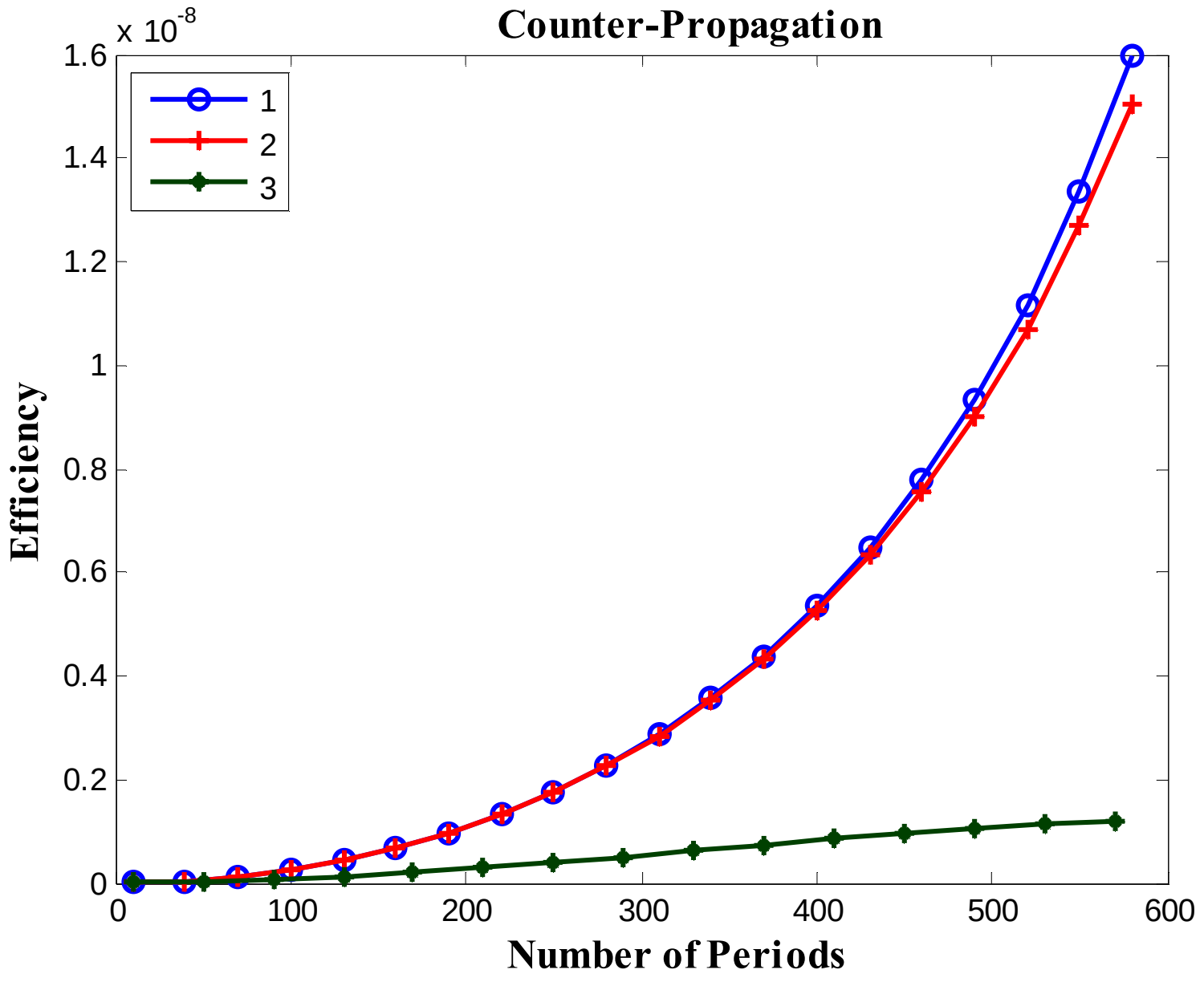

Fig. (6): Efficiency Vs. Number of Periods

$1)\,\delta = 0, 2)\,\delta = 0.001k_0, 3)\,\delta = 0.01k_0, n_0 = 3.45, \Delta k = 0, a_1 = 0.001, a_2 = -10^{-15}$

Effect of nonlinear index of refraction coefficient in different number of layers, on counter-propagation conversion efficiency is illustrated in Fig. (7).

Conversion efficiency is decreased with decreasing of the nonlinear index of refraction coefficient. FWM phenomenon is strongly affected by the nonlinear index of refraction coefficient and this effect has been shown in our numerical simulations.

Also, as we mentioned before, since the conversion efficiency is nonlinearly dependant on the number of periods, it is decreased when we choose much more layers than 600 layers, so it is necessary to keep in mind this subject in design of a photonic crystal to avoid decreasing of conversion efficiency.

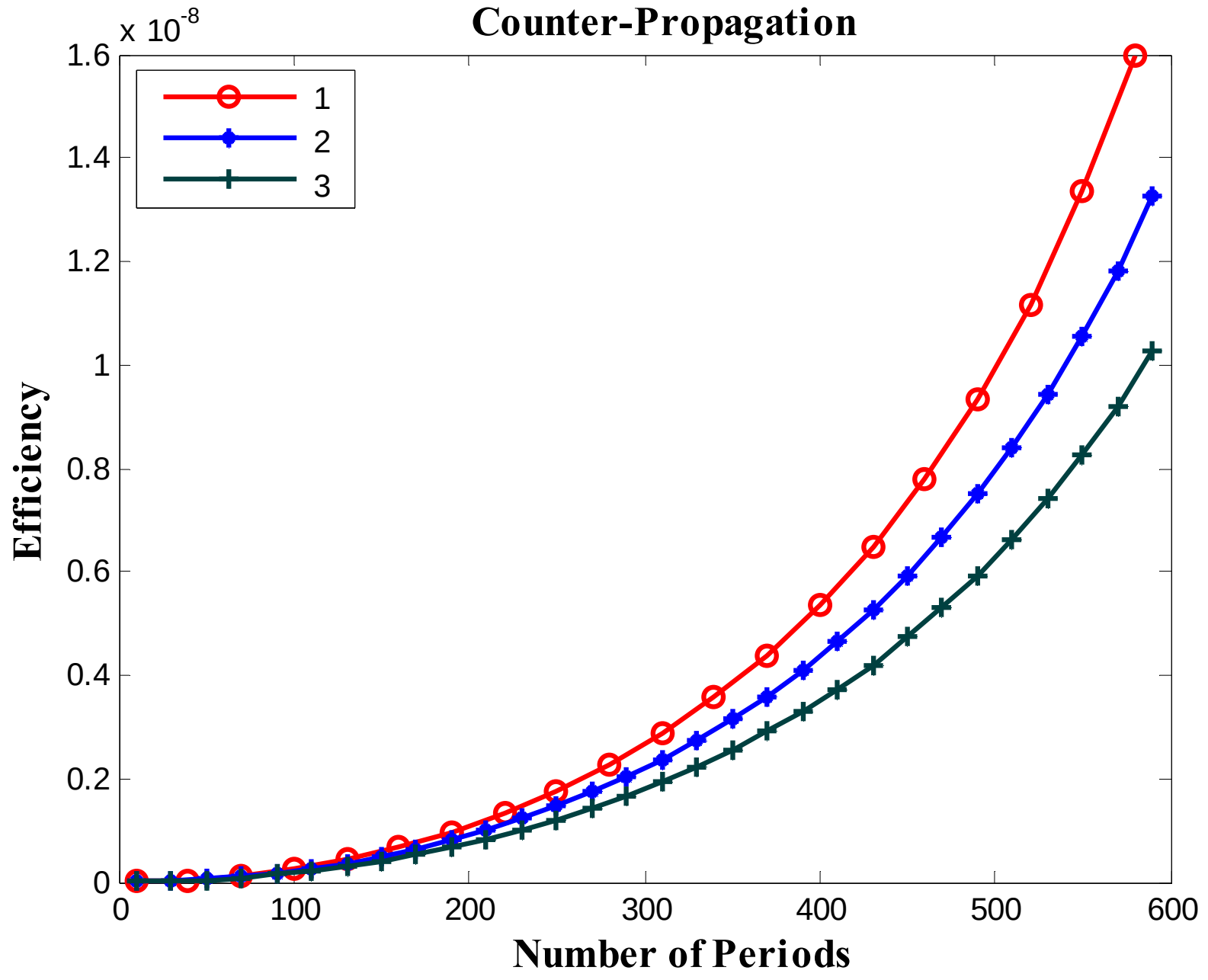

Fig. (7): Efficiency Vs. Number of Periods

$1)\, a_2 = -10^{-15}, 2)\, a_2 = -2\times10^{-15}, 3)\, a_2 = -3\times10^{-15}, n_0 = 3.45, \Delta k = 0, a_1 = 0.001, \delta = 0$

The effect of mismatching between average wave vector of four optical fields and Grating wave vector, is demonstrated in Fig. (8).

by increasing of mismatching, conversion efficiency decreases.

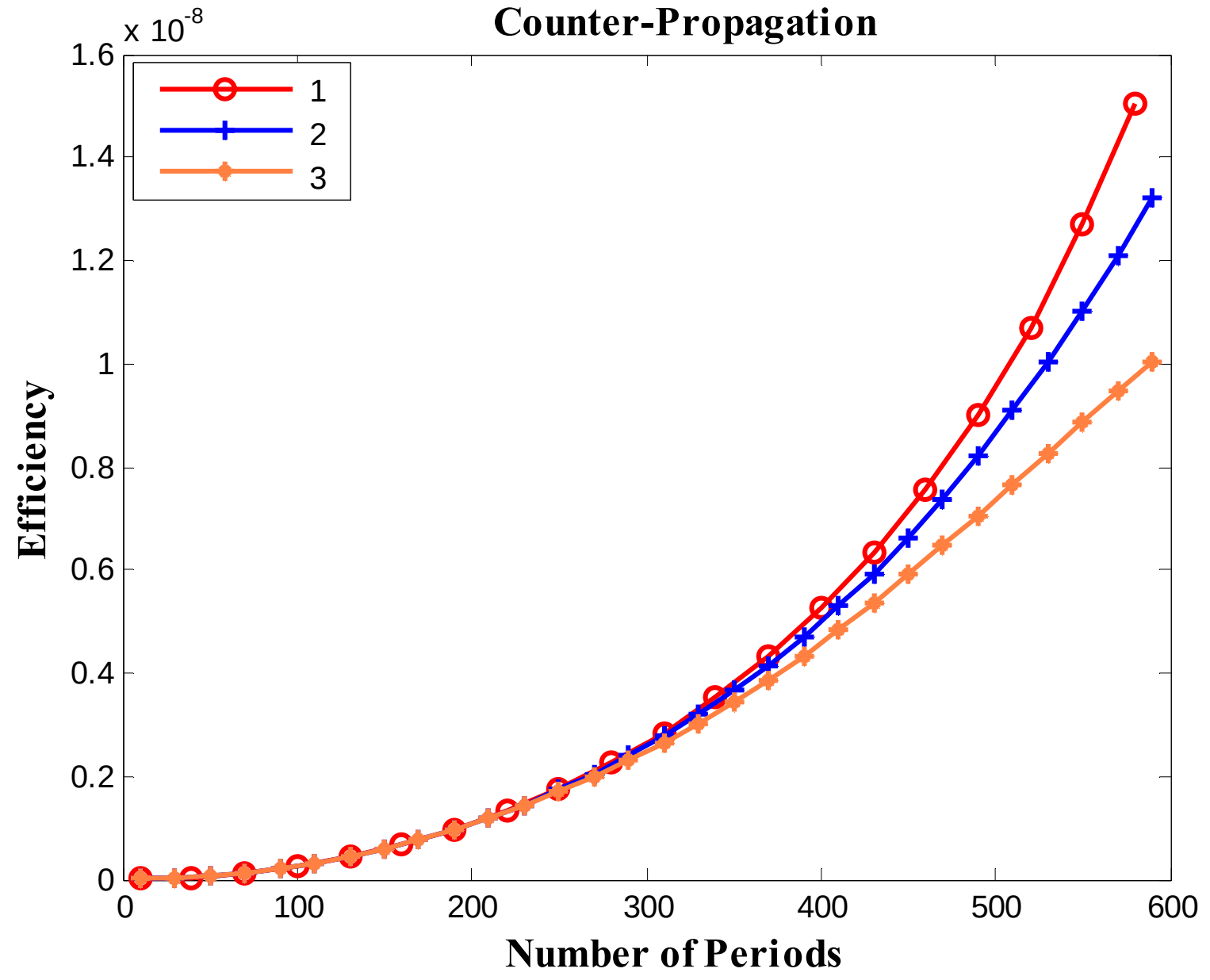

Fig. (8): Efficiency Vs. Number of Periods

$1)\, \delta = 0.001k_0, 2)\, \delta = 0.002k_0, 3)\, \delta = 0.003k_0, n_0 = 3.45, \Delta k = 0, a_1 = 0.001, a_2 = -10^{-15}$

The effect of phase mismatching between Grating and the average wave vectors, on the conversion efficiency in broad ranges is shown in Fig. (9).

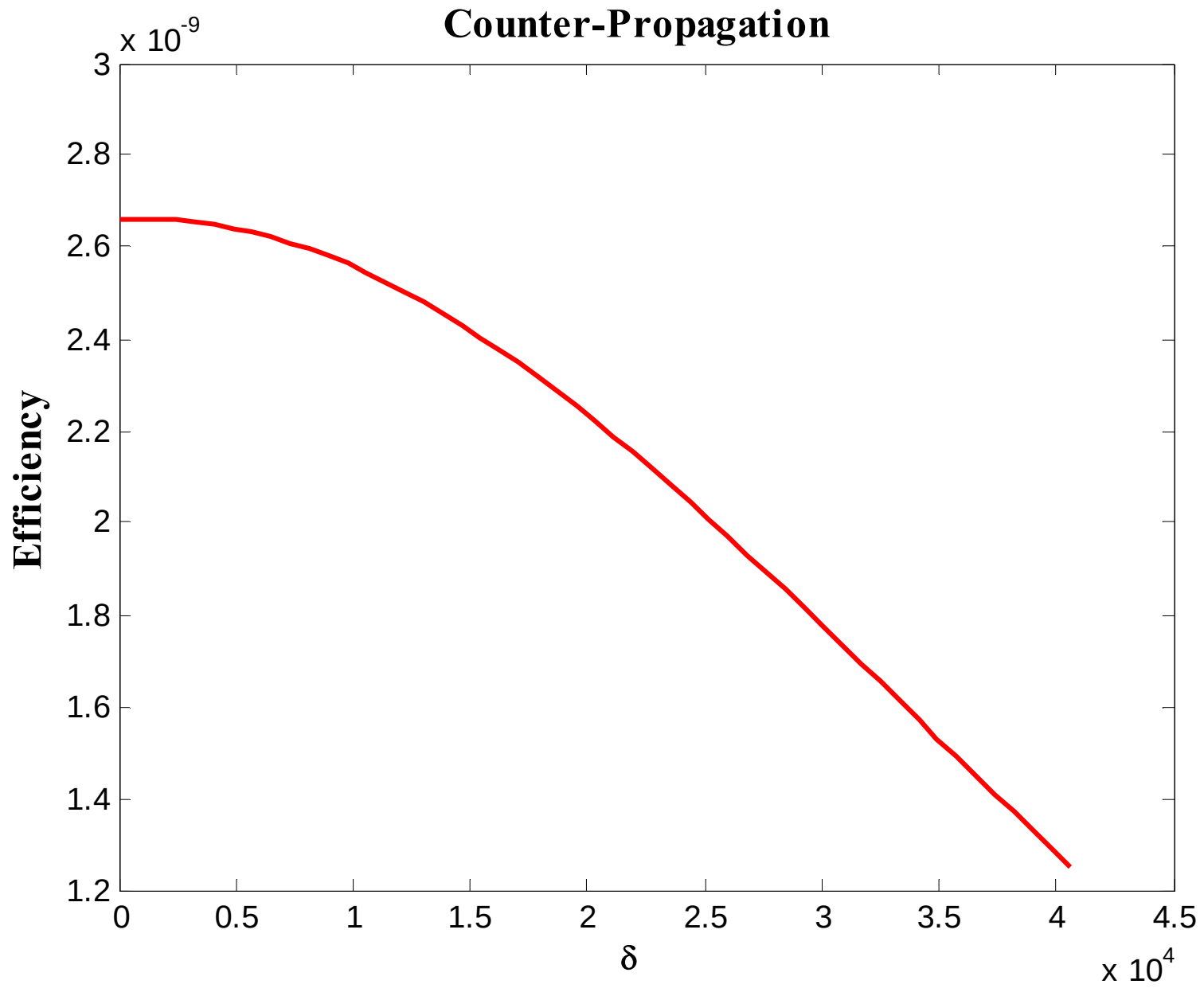

Fig. (9): Efficiency Vs. Phase mismatching between Medium and average applied wave vectors

$n_0 = 3.45, \Delta k = 0, a_1 = 0.001, a_2 = -10^{-15}, N = 300$

Also, the effect of the nonlinear index of refraction, on conversion efficiency is shown in Fig. (10).

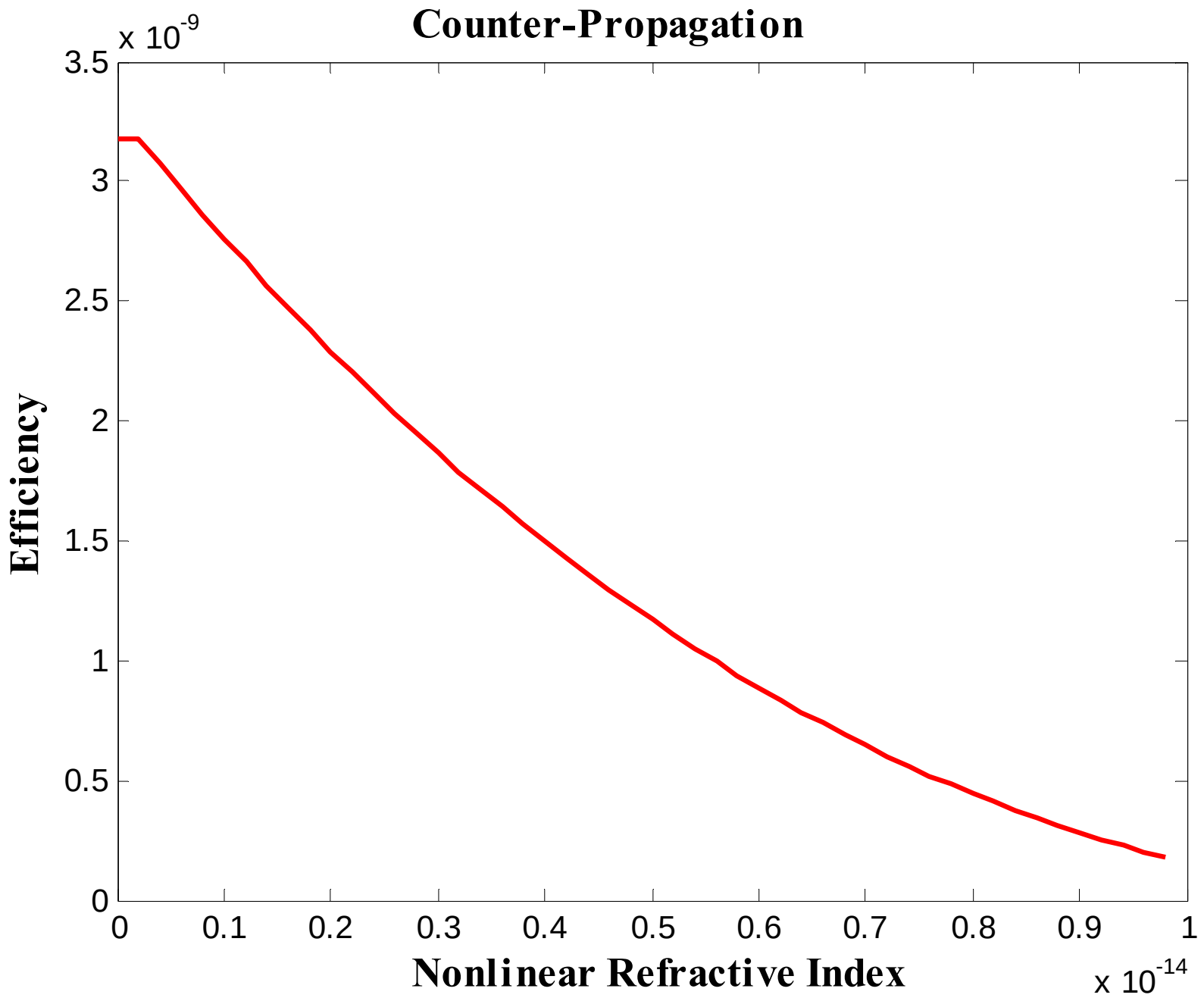

Fig. (10): Efficiency Vs. Absolute Value of Nonlinear Index of Refraction

$n_0 = 3.45, \Delta k = 0, a_1 = 0.001, \delta = 0, N = 300$

In this section, the numerical evaluation of the derived equations for light propagation through one-dimensional nonlinear photonic crystals was presented using FWM process. It was shown that by using nonlinear photonic crystal, conversion efficiency enhances.

**Summary and Conclusion-** The coupled mode equations for light propagation through one-dimensional nonlinear photonic crystals using FWM process in steady state condition have been developed. We have considered the photonic crystal as a lossless, dispersiveless and inhomogeneous medium. The linear and nonlinear indexes of refraction are approximated with the first Fourier harmonic term. The derived equations have been solved numerically and the obtained results show that conversion efficiency increases considerably using nonlinear photonic crystal. Also, the effect of system parameters on conversion efficiency was illustrated. It was shown that number of periods, nonlinear and linear indexes of refraction and magnitude of pump fields have a deep and strong effect on system performance. Although all of fields in our work deplete through propagating in nonlinear photonic crystal, we could enhance conversion efficiency of FWM process in these crystals. It should be mentioned that the presented results are not the optimum ones, but they are only typical simulations for some given parameters.